\pdfoutput=1
\documentclass[10pt,a4paper]{article}
\usepackage[margin=2cm]{geometry}
\usepackage{amsmath,amssymb}
\usepackage{graphicx}
\usepackage{float}
\usepackage{authblk}
\usepackage{hyperref}
\hypersetup{
  pdftitle={Flavor Democracy Calls for Vector Like Leptons and Quarks},
  pdfauthor={Burak Dagli, Saleh Sultansoy, Ismail Toy}
}

\title{Flavor Democracy Calls for Vector Like Leptons and Quarks}
\author[1]{Burak DA\u{G}LI}
\author[1]{Saleh SULTANSOY}
\author[1]{\.{I}smail TOY\thanks{Corresponding author: ismailtoy@etu.edu.tr}}
\affil[1]{TOBB University of Economics and Technology, Ankara, T\"urkiye}
\date{}

\begin{document}

\maketitle

\begin{abstract}
There are strong arguments favoring the Flavor Democracy hypothesis (or the Democratic Mass Matrix approach) within the Standard Model framework. However, the large mass of the top quark ($m_t >> m_b, m_\tau$) poses an obstacle to the functioning of Flavor Democracy in the three SM family scenario. While a fourth Standard Model generation could have provided a natural resolution, this possibility is now almost entirely excluded by precision data on Higgs boson production and decay rates. The Flavor Democracy hypothesis can be elegantly resurrected through the introduction of Vector-Like Leptons (VLLs) and Vector-Like Quarks (VLQs), which naturally accommodate the observed fermion mass hierarchies while remaining consistent with current experimental constraints. Currently, experimental searches for VLLs conducted by the ATLAS and CMS collaborations rely on a highly constrained Restricted Model. This model imposes a mass degeneracy between charged and neutral VLLs within a doublet and assumes the absence of right-handed neutrinos. Consequently, current results are valid only for the restricted model and do not cover a more realistic, general scenario. Therefore, to accurately reflect the physical reality, it is imperative to conduct a comprehensive re-evaluation that incorporates all viable decay channels.
\end{abstract}

\tableofcontents

\section{Introduction}

The validation of the Standard Model (SM) of particle physics culminated spectacularly with the discovery of the Higgs boson by the ATLAS and CMS collaborations \cite{ATLAS:2012,CMS:2012}. Despite its astonishing success as a low-energy effective field theory, the SM leaves several profound physical phenomena unexplained. Foremost among these is the "flavor puzzle," which questions the origins of the highly disparate mass and mixing patterns observed across the three generations of fundamental fermions. The late Steven Weinberg frequently cited this arbitrary pattern of quark and lepton masses as one of the preeminent unresolved mysteries in particle physics \cite{Chalmers:2017}, highlighting that the SM parameters have been dialed by hand to match experimental data (see Table 1). Furthermore, the mixing topologies between the quark and lepton sectors are structurally incompatible: the Cabibbo-Kobayashi-Maskawa (CKM) matrix exhibits a highly hierarchical structure, whereas the Pontecorvo-Maki-Nakagawa-Sakata (PMNS) matrix displays large, anarchic off-diagonal mixing angles.

\begin{table}[h]
\centering
\begin{tabular}{|c|c|c|c|}
\hline
\textbf{Generation} & \textbf{Charged Lepton ($l$)} & \textbf{Up Type Quark ($u$)} & \textbf{Down Type Quark ($d$)} \\
\hline
1 & $m_e \approx 0.511$ MeV & $m_u \approx 2.2$ MeV & $m_d \approx 4.7$ MeV \\
\hline
2 & $m_\mu \approx 105.7$ MeV & $m_c \approx 1.27$ GeV & $m_s \approx 96$ MeV \\
\hline
3 & $m_\tau \approx 1.777$ GeV & {\boldmath\bfseries $m_t \approx 172.7$ GeV} & $m_b \approx 4.18$ GeV \\
\hline
\end{tabular}
\caption{Charged Fermion Masses by Generation}
\end{table}

The Flavor Democracy (FD) hypothesis provides a natural and compelling resolution to this problem (see for example \cite{Sultansoy:2007} and references therein). Formulated initially for a three-family SM, the strict democratic mass matrix stipulates that all fermions of a given charge interact with the Higgs field with equal strength before small symmetry-breaking perturbations are introduced. While a strict three-generation democracy was ruled out by the overwhelming mass of the top quark and the fourth generation is excluded by precision data on Higgs boson production and decay rates, the hypothesis is beautifully resurrected by the introduction of Vector-Like Fermions (VLFs) \cite{Baspehlivan:2022}. Unlike standard chiral fermions, the left and right-handed components of VLFs transform identically under the SM gauge group, allowing them to acquire bare Dirac masses without relying exclusively on the Higgs mechanism, thereby safely decoupling from precision electroweak bounds \cite{Acar:2021}.

Crucially, the necessity of VLFs is not merely a bottom-up phenomenological convenience, namely FD; it is an explicit mathematical prediction of Grand Unified Theory (GUT) based on the $E_6$ gauge group \cite{Gursey:1976}, which is strongly favored by superstring theory (see \cite{Hewett:1989} and references therein). The decomposition of the fundamental 27-dimensional representation of $E_6$ mandates the existence of one iso-singlet down-type quark and an iso-doublet of vector-like leptons (VLLs) per SM generation. Moreover, beyond structural elegance, vector-like leptons and quarks can provide solutions to some of the anomalies we encounter in particle physics \cite{Crivellin:2024}.

Despite possessing equal phenomenological status, the experimental search landscape exhibits a profound and unjustified bias favoring vector-like quarks (VLQs) over vector-like leptons. The ATLAS and CMS collaborations have conducted extensive searches for VLQs, while searches for VLLs remain severely restricted. Historical bounds from the LEP L3 collaboration only excluded VLLs up to approximately 100 GeV \cite{L3:2001}. Current LHC searches have been narrowed almost entirely to third-generation partners (coupled to the tau lepton) \cite{ATLAS:2023vll,CMS:2022}, erroneously importing the hierarchical assumptions of the CKM matrix into the inherently anarchic lepton sector \cite{Baspehlivan:2022}.

More critically, current VLL searches rely on a highly constrained "Restricted Model" (RM) that assumes mass degeneracy between charged and neutral VLLs in a doublet, and assuming absence of right-handed neutrino components \cite{Bhattiprolu:2019}. This eliminates several important decay channels, such as the neutral current decay $N \to Z\nu$ and inter-multiplet cascade decays like $E \to WN$, predicted by more realistic "General Model" (GM) \cite{Baspehlivan:2023}. Therefore, it is imperative that future search strategies at the High-Luminosity LHC and proposed energy-frontier lepton and hadron colliders expand to encompass first-and second-generation vector-like leptons, integrating a complete array of topological signatures to fully probe this promising sector of new physics \cite{Baspehlivan:2023}.

This article is structured as follows. In Section 2, we present the basic assumptions of the Flavor Democracy Hypothesis, its application to the Standard Model, and the challenges encountered during this application. In Section 3, we demonstrate that the inclusion of vector-like leptons and quarks solves these challenges. Production of vector-like leptons at the LHC has been discussed in Section 4, highlighting processes that current analyses have overlooked. Finally, we conclude in Section 5.

\section{Flavor Democracy and Standard Model}

\subsection{Higgs Mechanism and Mass Eigenvalues}

It is useful to consider two different bases \cite{Ciftci:2005}:

(i) Standard model basis $\{f^0\}$ before spontaneous symmetry breaking (SSB)

(ii) Mass basis $\{f\}$ after SSB.

According to the three-family SM, before the spontaneous symmetry breaking quarks are grouped into the following $SU(2)\times U(1)$ multiplets,
\begin{equation}
\begin{pmatrix} u^0_L \\ d^0_L \end{pmatrix}, u^0_R, d^0_R; \quad
\begin{pmatrix} c^0_L \\ s^0_L \end{pmatrix}, c^0_R, s^0_R; \quad
\begin{pmatrix} t^0_L \\ b^0_L \end{pmatrix}, t^0_R, b^0_R.
\end{equation}
In the one-family case two bases are equal and, for example, d-quark mass is obtained due to
\begin{equation}
L_Y^{(d)} = a_d(\overline{u}^0_L \quad \overline{d}^0_L)\binom{\varphi^+}{\varphi^0} d^0_R + H.c. \Rightarrow L_m^{(d)} = m_d\overline{d}d,
\end{equation}
where $m_d = a_d\eta,$ $\eta = \langle\varphi^0\rangle \cong$ 249 GeV. In the same manner $m_u = a_u\eta,$ $m_e = a_e\eta,$ and $m_{\nu_e} = a_{\nu_e}\eta$ (if the neutrino is a Dirac particle).

In the n-family case,
\begin{eqnarray}
L_Y^{(d)} &=& \sum_{i,j=1}^{n} a^d_{ij}(\overline{u^0}_{Li} \quad \overline{d^0}_{Li})\binom{\varphi^+}{\varphi^0} d^0_{Rj} + H.c. \nonumber \\
&\Rightarrow& L_m^{(d)} = \sum_{i,j=1}^{n} m^d_{ij}\overline{d^0}_i d^0_j, \quad m^d_{ij} = a^d_{ij}\eta,
\end{eqnarray}
where $d^0_1$ denotes $d^0$, $d^0_2$ denotes $s^0$, etc. The diagonalization of the mass matrix of each type of fermion, or in other words transition from SM basis to mass basis, is performed by well-known bi-unitary transformation:
\begin{equation}
d_{iL} = (U^d_L)_{ij}d^0_{jL}, \quad d_{iR} = (U^d_R)_{ij}d^0_{jR}
\end{equation}
similarly,
\begin{align}
u_{iL} &= (U^u_L)_{ij}u^0_{jL}, \quad u_{iR} = (U^u_R)_{ij}u^0_{jR}, \nonumber \\
l_{iL} &= (U^l_L)_{ij}l^0_{jL}, \quad l_{iR} = (U^l_R)_{ij}l^0_{jR}, \\
\nu_{iL} &= (U^\nu_L)_{ij}\nu^0_{jL}, \quad \nu_{iR} = (U^\nu_R)_{ij}\nu^0_{jR} \nonumber
\end{align}
The last expression is valid for the Dirac neutrino, while the situation is more complicated for Majorana neutrinos.

If one takes only electromagnetic interactions into consideration, one gets
\begin{equation}
J^0_{em}(d) = q_d\sum_i(\overline{d^0_{iL}}\gamma_\mu d^0_{iL} + \overline{d^0_{iR}}\gamma_\mu d^0_{iR})
\end{equation}
in the SM basis. When the transformation from SM basis to mass basis is performed with the use of inverse of Eq. (4), one obtains
\begin{equation}
J_{em}(d) = q_d\sum_{i,j,k}(\overline{d_{kL}}(U^d_L)_{ki}\gamma_\mu(U^{d^\dagger}_L)_{ij}d_{jL} + \overline{d_{kR}}(U^d_R)_{ki}\gamma_\mu(U^{d^\dagger}_R)_{ij}d_{jR}).
\end{equation}
Since
\begin{equation}
\begin{aligned}
\sum_i(U^d_L)_{ki}(U^{d^\dagger}_L)_{ij} &= (U^d_L U^{d^\dagger}_L)_{kj} = \delta_{kj}, \\
\sum_i(U^d_R)_{ki}(U^{d^\dagger}_R)_{ij} &= (U^d_R U^{d^\dagger}_R)_{kj} = \delta_{kj}
\end{aligned}
\end{equation}
one obtains
\begin{equation}
J_{em}(d) = q_d\sum_k(\overline{d_{kL}}\gamma_\mu d_{kL} + \overline{d_{kR}}\gamma_\mu d_{kR}).
\end{equation}
As one can observe, the electromagnetic current is not changed with transformation from SM to mass basis. A similar situation takes place for interactions with the Z boson.

In the case of charged weak current,
\begin{equation}
J^0_W = \frac{g}{\sqrt{2}}\sum_i\overline{u^0_{iL}}\gamma_\mu d^0_{iL}
\end{equation}
in the SM basis. The transformation from SM basis to mass basis leads to
\begin{equation}
J_W = \frac{g}{\sqrt{2}}\sum_{i,j,k}\overline{u_{kL}}(U^u_L)_{ki}\gamma_\mu(U^{d^\dagger}_L)_{ij}d_{jL},
\end{equation}
where
\begin{equation}
\sum_i(U^u_L)_{ki}(U^{d^\dagger}_L)_{ij} = (U^u_L U^{d^\dagger}_L)_{kj} \neq \delta_{kj}.
\end{equation}
In this context the well-known CKM matrix is defined as
\begin{equation}
U_{CKM} = U^u_L(U^d_L)^\dagger
\end{equation}

\subsection{Flavor Democracy Hypothesis}

Before the spontaneous symmetry breaking, all quarks are massless and there are no differences between $d^0$, $s^0$ and $b^0$. In other words, fermions with the same quantum numbers are indistinguishable. This leads us to \textit{the first assumption} \cite{Harari:1978,Fritzsch:1987,Fritzsch:1990,Fritzsch:1979,Kaus:1988}; namely, Yukawa couplings are equal within each type of fermion:
\begin{equation}
a^d_{ij} = a^d, \quad a^u_{ij} = a^u, \quad a^l_{ij} = a^l, \quad a^\nu_{ij} = a^\nu.
\end{equation}
In the case of $n$ SM families, the first assumption results in $n-1$ massless particles and one massive particle with $m = na^F\eta$ ($F = u,d,l,\nu$) for each type of the SM fermion. Comparing the values in Table 1, it appears that the first assumption solves the hierarchy problem for charged leptons, up-type quarks, and down-type quarks. The mass values of the first two families of fermions can be obtained thanks to small deviations from flavor democracy.

Because there is only one Higgs doublet which gives Dirac masses to all four types of fermions, it seems natural to make \textit{the second assumption} \cite{Datta:1994,Celikel:1995}; namely, the Yukawa constant for different types of fermions should naturally be equal:
\begin{equation}
a^d = a^u = a^l = a^\nu.
\end{equation}

\subsection{Three SM families}

For the 3 SM family case, in terms of the mass matrix, the above arguments mean
\begin{equation}
M^0 = a\eta\begin{bmatrix} 1 & 1 & 1 \\ 1 & 1 & 1 \\ 1 & 1 & 1 \end{bmatrix} \Rightarrow M = a\eta\begin{bmatrix} 0 & 0 & 0 \\ 0 & 0 & 0 \\ 0 & 0 & 3 \end{bmatrix}
\end{equation}
resulting in $m_b = m_\tau = m_t = 3a\eta$.

Taking into account the mass values for the third generation ($m_b \sim m_\tau \ll~ m_t = 175 GeV$), the second assumption leads to the statement that, according to the flavor democracy, the fourth SM family should exist \cite{Celikel:1995,Fritzsch:1992,Datta:1993}.

\subsection{The Fourth SM family}

In terms of the mass matrix, the above arguments mean
\begin{equation}
M^0 = a\eta\begin{bmatrix} 1 & 1 & 1 & 1 \\ 1 & 1 & 1 & 1 \\ 1 & 1 & 1 & 1 \\ 1 & 1 & 1 & 1 \end{bmatrix} \Rightarrow M = a\eta\begin{bmatrix} 0 & 0 & 0 & 0 \\ 0 & 0 & 0 & 0 \\ 0 & 0 & 0 & 0 \\ 0 & 0 & 0 & 4 \end{bmatrix}
\end{equation}
resulting in $m_{d4} = m_{e4} = m_{u4} = 4a\eta$. In order to give nonzero masses for the first three SM family fermions, flavor democracy has to be slightly broken \cite{Ciftci:2005,Atag:1996}. However, the 4th SM family is almost excluded by recent data on the Higgs production cross-section and decay rates \cite{Eberhardt:2012,Djouadi:2012}, since in case of a heavier extra family the rate of Higgs production by gluon fusion would have been much more than have been seen by the LHC.

\section{Solution to The Problem: Vector Like Leptons and Quarks}

This discrepancy can be solved by the introduction of vector-like leptons and quarks to the 3 family SM (VLQ's did not enter to ggH loop since their masses do not come from Higgs mechanism).

\subsection{Only One Vector-like Family}

The mass pattern of the SM fermions may be enlightened by the flavor democracy (democratic mass matrix) hypothesis through inclusion of vector-like leptons and quarks. Recently this possibility is reconsidered in \cite{Acar:2021,Kaya:2019}. In this section, we briefly summarize corresponding part of \cite{Kaya:2019}.

\textbf{a) Iso-singlet quark}

The quark sector of SM is modified by addition of iso-singlet down-type quark Q:
\begin{equation}
\begin{pmatrix} u \\ d \end{pmatrix}_L, \begin{pmatrix} c \\ s \end{pmatrix}_L, \begin{pmatrix} t \\ b \end{pmatrix}_L, u_R, c_R, t_R, d_R, s_R, b_R, Q_L, Q_R
\end{equation}

Let us note that it is unnatural for up-type quarks to have vector-like partners under the flavor democracy hypothesis, because of high mass of t-quark ($m_t \simeq 175$ GeV, which is close to vacuum expectation value of the Higgs field), whereas the addition of the vector-like partners of the down-type quarks explains why the b-quark has a much lower mass than the t-quark. In the case of full Flavor Democracy, the mass matrix of the up-type quarks can be written as
\begin{equation}
M_u = \begin{pmatrix} a\eta & a\eta & a\eta \\ a\eta & a\eta & a\eta \\ a\eta & a\eta & a\eta \end{pmatrix}
\end{equation}
and mass matrix of down type quarks is
\begin{equation}
M_d = \begin{pmatrix} a\eta & a\eta & a\eta & a\eta \\ a\eta & a\eta & a\eta & a\eta \\ a\eta & a\eta & a\eta & a\eta \\ M & M & M & M \end{pmatrix}
\end{equation}
where $\eta = 246$ GeV is vacuum expectation value of Higgs field and $M$ ($M >> \eta$) is the new physics scale that determines the mass of iso-singlet quark. In this case $m_u = m_c = 0$ and $m_t = 3a\eta$ for up type quarks, $m_d = m_s = m_b = 0$ and $m_Q = 3a\eta + M = m_t + M$ for down type quarks.

The masses of lighter quarks can be obtained through slight modifications of flavor democracy \cite{Acar:2021,Kaya:2019}.

\textbf{b) Vector-like leptons}

In a similar manner low value of $\tau$ lepton mass can be provided by adding an iso-singlet charged lepton or iso-doublet vector-like leptons. In iso-doublet case, lepton multiples are given below:
\begin{equation}
\begin{pmatrix} \nu_e \\ e \end{pmatrix}_L, \begin{pmatrix} \nu_\mu \\ \mu \end{pmatrix}_L, \begin{pmatrix} \nu_\tau \\ \tau \end{pmatrix}_L, \nu_{eR}, \nu_{\mu R}, \nu_{\tau R}, e_R, \mu_R, \begin{pmatrix} L^0 \\ L^- \end{pmatrix}_L, \begin{pmatrix} L^0 \\ L^- \end{pmatrix}_R
\end{equation}

Let us emphasize that right-handed neutrinos should be included into SM since (according to lepton-quark symmetry) they are counterparts of right-handed components of up quarks. This statement is confirmed by the observation of neutrino oscillations. Earlier, absence of $\nu_R$ was postulated mistakenly because of V-A structure of charged weak currents.

Masses of $\tau$-lepton and muon are generated by following modification of the charged lepton mass matrix:
\begin{equation}
M_l = \begin{pmatrix}
a\eta & a\eta & a\eta & (1-\beta_\tau)M \\
a\eta & a\eta & a\eta & (1-\beta_\tau)M \\
a\eta & a\eta & (1+\alpha_\mu)a\eta & (1-\beta_\tau)M \\
(1-\alpha_\tau)a\eta & (1-\alpha_\tau)a\eta & (1-\alpha_\tau)a\eta & M
\end{pmatrix}
\end{equation}
For $M = 2000$ GeV, $\alpha_\tau = \beta_\tau = 5.58\times10^{-3}$ and $\alpha_\mu = 2.73\times10^{-4}$ this mass matrix leads to $m_L = 2171$ GeV, $m_\tau = 1.777$ GeV, $m_\mu = 104.7$ MeV, $m_e = 0$.

Zero masses for the SM neutrinos can be provided by introduction iso-singlet vector-like heavy neutral lepton $N_L$ and $N_R$. Corresponding mass matrix has a form:
\begin{equation}
M_\nu = \begin{pmatrix} a\eta & a\eta & a\eta & M \\ a\eta & a\eta & a\eta & M \\ a\eta & a\eta & a\eta & M \\ a\eta & a\eta & a\eta & M \end{pmatrix}
\end{equation}
Diagonalization of this matrix leads to $m(\nu_e) = m(\nu_\mu) = m(\nu_\tau) = 0$ and $m_N = M + m_t$.

\subsection{Three Vector-like Families}

It is more natural for each standard model family to have it's corresponding vector-like quarks and leptons. In this case, quark sector becomes:
\begin{equation}
\begin{pmatrix} u \\ d \end{pmatrix}_L, \begin{pmatrix} c \\ s \end{pmatrix}_L, \begin{pmatrix} t \\ b \end{pmatrix}_L, u_R, c_R, t_R, d_R, s_R, b_R, D_L, D_R, S_L, S_R, B_L, B_R
\end{equation}
and lepton sector is modified as:
\begin{equation}
\begin{pmatrix} \nu_e \\ e \end{pmatrix}_L, \begin{pmatrix} \nu_\mu \\ \mu \end{pmatrix}_L, \begin{pmatrix} \nu_\tau \\ \tau \end{pmatrix}_L, \nu_{eR}, \nu_{\mu R}, \nu_{\tau R}, e_R, \mu_R, \begin{pmatrix} N_e \\ E^- \end{pmatrix}_L, \begin{pmatrix} N_e \\ E^- \end{pmatrix}_R, \begin{pmatrix} N_\mu \\ M^- \end{pmatrix}_L, \begin{pmatrix} N_\mu \\ M^- \end{pmatrix}_R, \begin{pmatrix} N_\tau \\ T^- \end{pmatrix}_L, \begin{pmatrix} N_\tau \\ T^- \end{pmatrix}_R
\end{equation}

Therefore, the mass matrix for bottom quarks takes the following form:

\begin{equation}
M_d = \begin{pmatrix}
a\eta & a\eta & a\eta & a\eta & a\eta & a\eta \\
a\eta & a\eta & a\eta & a\eta & a\eta & a\eta \\
a\eta & a\eta & a\eta & a\eta & a\eta & a\eta \\
M & M & M & M & M & M \\
M & M & M & M & M & M \\
M & M & M & M & M & M
\end{pmatrix}
\end{equation}

Similar modifications occur for leptons as well.

\subsection{E6 GUT Anticipates the Existence of VLLs and VLQs}

The first family fermion sector of the $E_6$-induced model has the following $SU_c(3)\times SU_w(2)\times U_Y(1)$ structure:
\begin{equation}
\begin{pmatrix} u_L \\ d_L \end{pmatrix} u_R\ d_R\ D_L\ D_R \begin{pmatrix} \nu_{eL} \\ e_L \end{pmatrix} \nu_{eR}\ e_R \begin{pmatrix} N_{eL} \\ E_L \end{pmatrix} \begin{pmatrix} N_{eL} \\ E_R \end{pmatrix} \mathcal{N}_e
\end{equation}

Similar structure is a case for second and third family fermions. Therefore, quark sector of the SM is extended by addition of three new iso-singlet down-type quarks. As for lepton sector, there are three new charged leptons, three new neutral Dirac leptons and three neutral Majorana leptons.

Hereafter we assume that interfamily mixings are dominant which means that new leptons will decay into their SM partners.

\section{Vector-like Leptons at the LHC}

The production and decays of vector-like leptons in the general model are discussed in detail in \cite{Baspehlivan:2022,Acar:2021,Baspehlivan:2023}. Let us reiterate that the restricted model is based on two incorrect assumptions: mass degeneracy between charged and neutral VLLs in a doublet(this assumption may be correct if there is only one isodoublet of vector like leptons), and assuming absence of right-handed neutrino components (this assumption has been ruled out due to neutrino oscillations.)

If the mass hierarchy of standard model leptons also holds for vector-like leptons, then it is highly probable that neutral VLLs are lighter than charged ones. For this reason, we consider the case where $N_e$ is the lightest VLL as an example.

\subsection{General Remarks on VLL Production and Decays}

Cross-sections for pair and associate productions of iso-doublet vector-like leptons at the LHC are shown in Figure 1, where cross-sections for associate E and N productions are calculated assuming $m_E = m_N$. It is seen that the $E^+N$ cross-section exceeds the other three approximately by a factor of 4. Since the masses are assumed to be equal, this figure applies to both RM and GM.

\begin{figure}[H]
\centering
\includegraphics[width=0.9\textwidth]{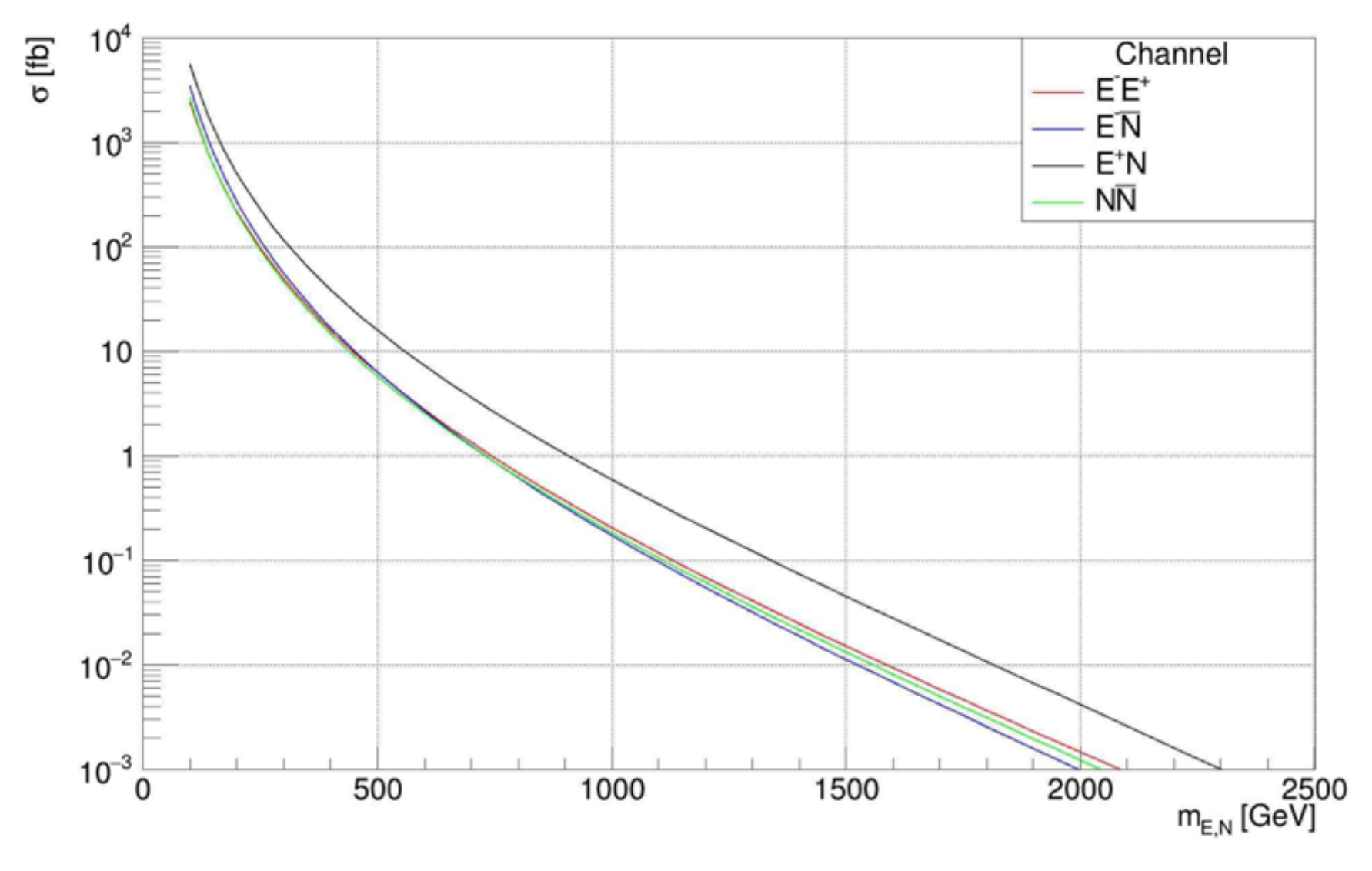}
\caption{Cross-sections for pair production of VLLs at the LHC with $\sqrt{s} = 14$ TeV in the case of $m_N = m_E$.}
\end{figure}

In Figure 2, we show cross-sections for the associated production of VLLs for different ratios of charged and neutral VLL masses. It is seen that the cross-section decreases as the ratio increases. Therefore, ATLAS and CMS exclusion limits are not applicable to non-degenerate case.

\begin{figure}[H]
\centering
\includegraphics[width=0.8\textwidth]{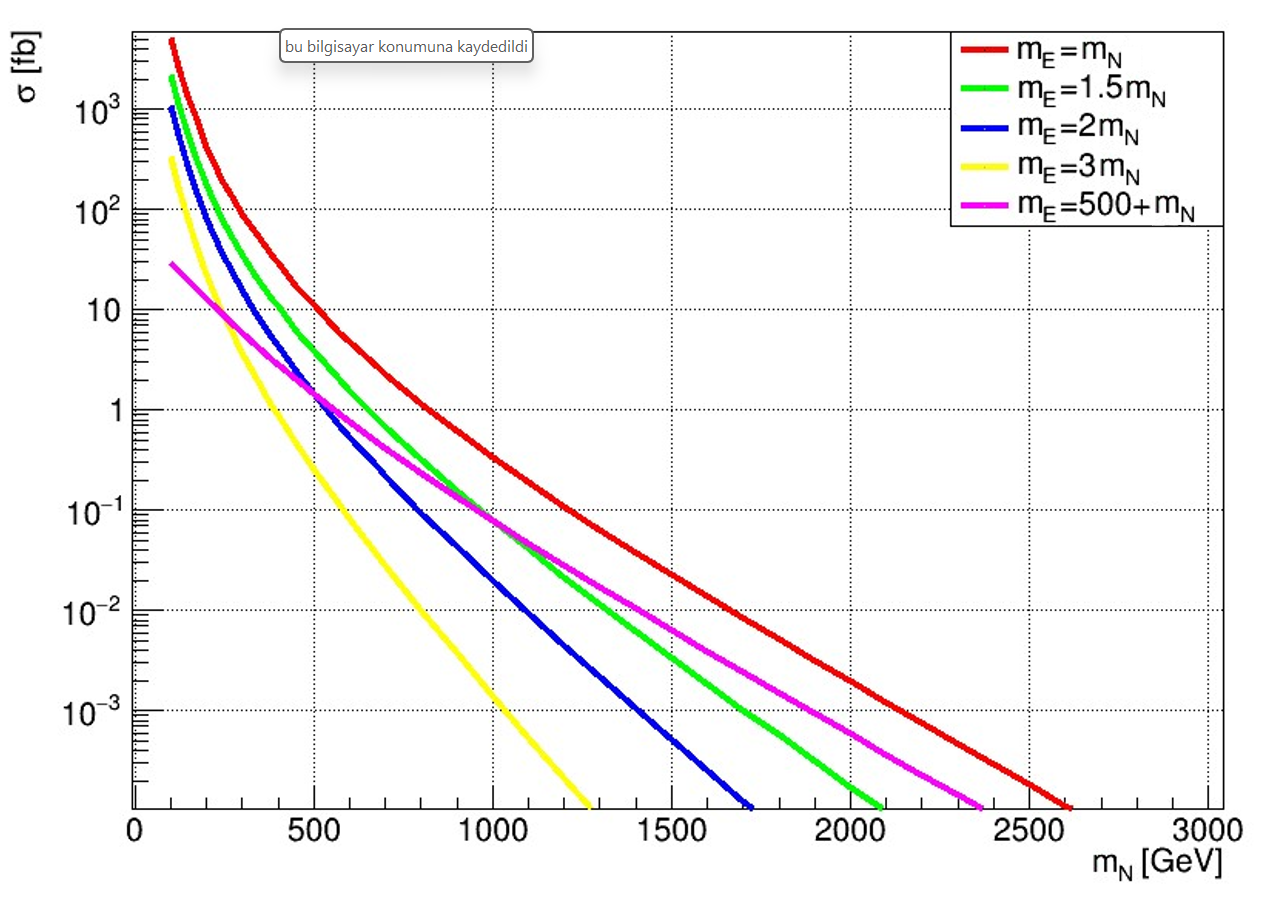}
\caption{Cross-sections for the associated production of VLLs for different ratios of charged and neutral VLL masses.}
\end{figure}

\vspace{\baselineskip}
Regarding vector-like lepton decays, the inclusion of right-handed neutrinos introduces new decay channels, $E^- \to W^-\nu$, $N \to H\nu$, and $N \to Z\nu$, in addition to the restricted model channels $E \to Ze$, $E \to He$, and $N \to We$. There are also new decay modes caused by different masses of charged and neutral VLLs, for example, $E^- \to W^-N$. If $m_E > m_N + m_W$, this unsuppressed mode becomes dominant, since widths of other decay modes are supressed by the square of sinus of the mixing angle. Corresponding Feynman diagrams are shown in Figure 3.

\begin{figure}[H]
\centering
\includegraphics[width=0.6\textwidth]{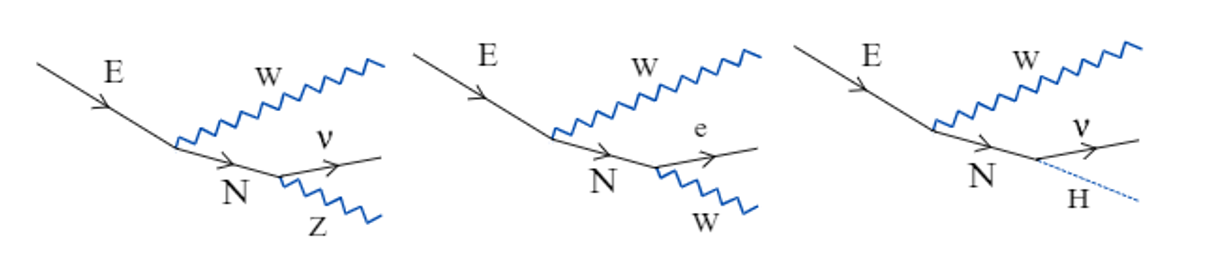}
\caption{Feynman diagrams for $m_E > m_N + m_W$ case}
\end{figure}

If this scenario occurs, the analyses performed will need to be completely revised.

\subsection{If \texorpdfstring{$N_e$}{Ne} is the Lightest VLL}

Decay width formulas for iso-doublet $N$ are (for details and notations, see \cite{Baspehlivan:2022}):
\begin{align}
\Gamma(N \to W^+e^-) &= \frac{m_N}{32\pi}a_N\left[(c^N_L s^E_L - c^E_L s^N_L)^2 + (c^N_R)^2(s^E_R)^2\right](1 - r^N_W)^2(1 + 2r^N_W) \\
\Gamma(N \to Z\nu) &= \frac{m_N}{64\pi}a_N(c^N_R)^2(s^N_R)^2(1 - r^N_Z)^2(1 + 2r^N_Z) \\
\Gamma(N \to H\nu) &= \frac{m_N}{64\pi}a_N(c^N_R)^2(s^N_R)^2(1 - r^N_H)^2
\end{align}

In the $m_N \gg m_{W,Z}$ case branching ratios become $\mathrm{BR}(N \to W^+e^-) = 0.5$ and $\mathrm{BR}(N \to Z\nu) = \mathrm{BR}(N \to H\nu) = 0.25$ if $s^E_L = s^N_L$ and $s^E_R = s^N_R$. Dependence of branching ratios on the mass of neutral vector-like lepton is presented in Figure 4.

\begin{figure}[H]
\centering
\includegraphics[width=0.6\textwidth]{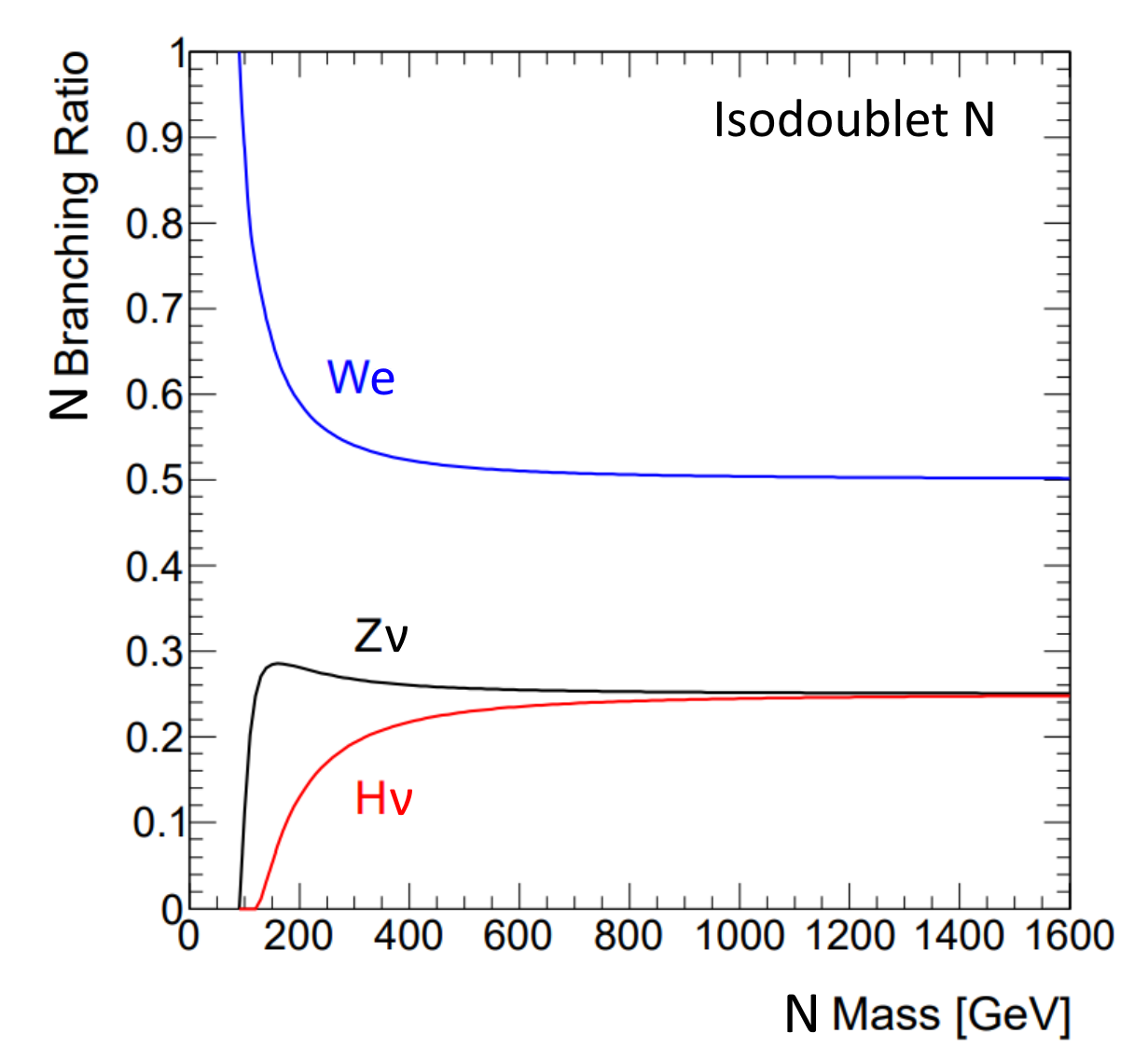}
\caption{Branching ratios for decays of iso-doublet neutral vector-like lepton ($s^E_L = s^N_L$ and $s^E_R = s^N_R$).}
\end{figure}

However, in the case of $s^E_L = s^N_L$ and $s^E_R = 0$, only neutral current decay modes are survived. Dependence of corresponding branching ratios on $m_N$ is presented in Figure 5. Similar situation take place if $s^N_R$ is dominant ($s^N_R \gg s^E_L, s^N_L, s^E_R$), since charged current decay mode is suppressed.

\begin{figure}[H]
\centering
\includegraphics[width=0.6\textwidth]{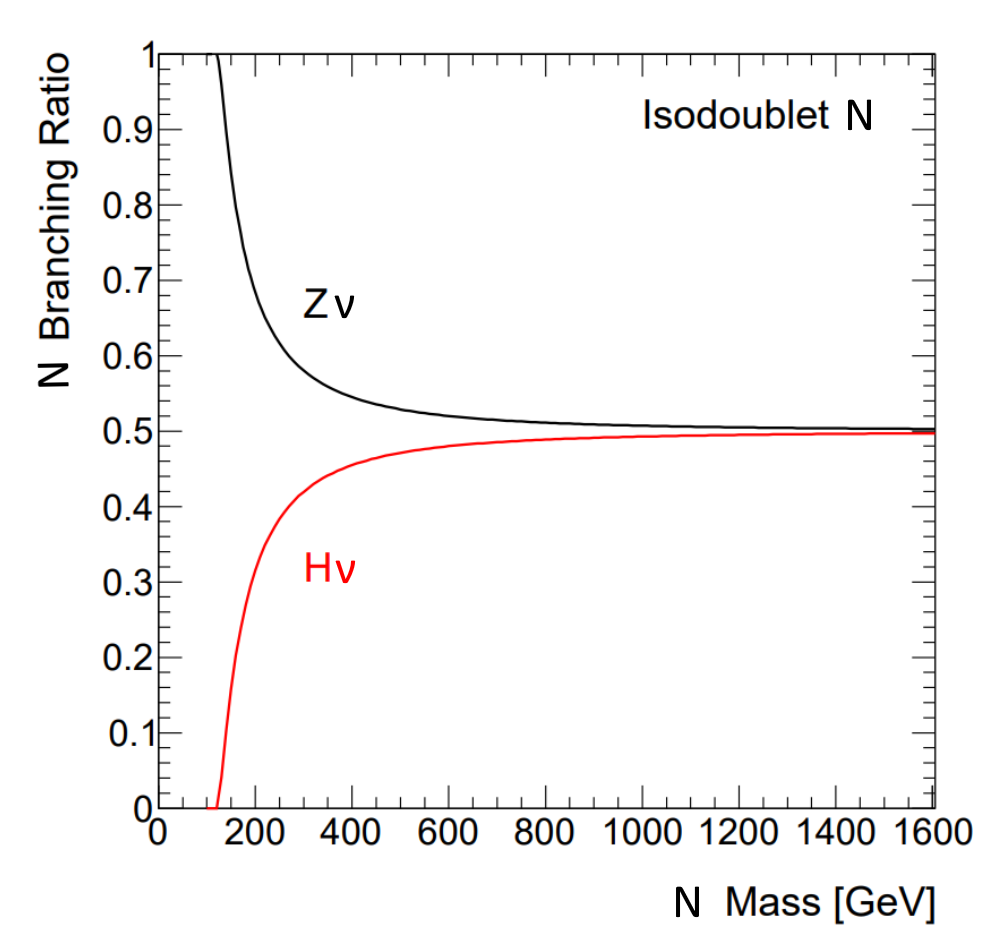}
\caption{ Branching ratios for decays of iso-doublet neutral vector-like lepton (neutral currents only)}
\end{figure}

The dependence of the decay widths on the neutral VLL mass is shown in Figure 6 in the case of only neutral current decay modes survived and $s^N_R = 0.01$. Let us remind that the decay width is proportional to $(s^N_R)^2$ for all masses. The decay width become proportional to $m^3_N$ for values exceeding to 1 TeV. It is seen that the decay width value is under experimental resolution even at $m_N = 5000$ GeV.

\begin{figure}[H]
\centering
\includegraphics[width=0.8\textwidth]{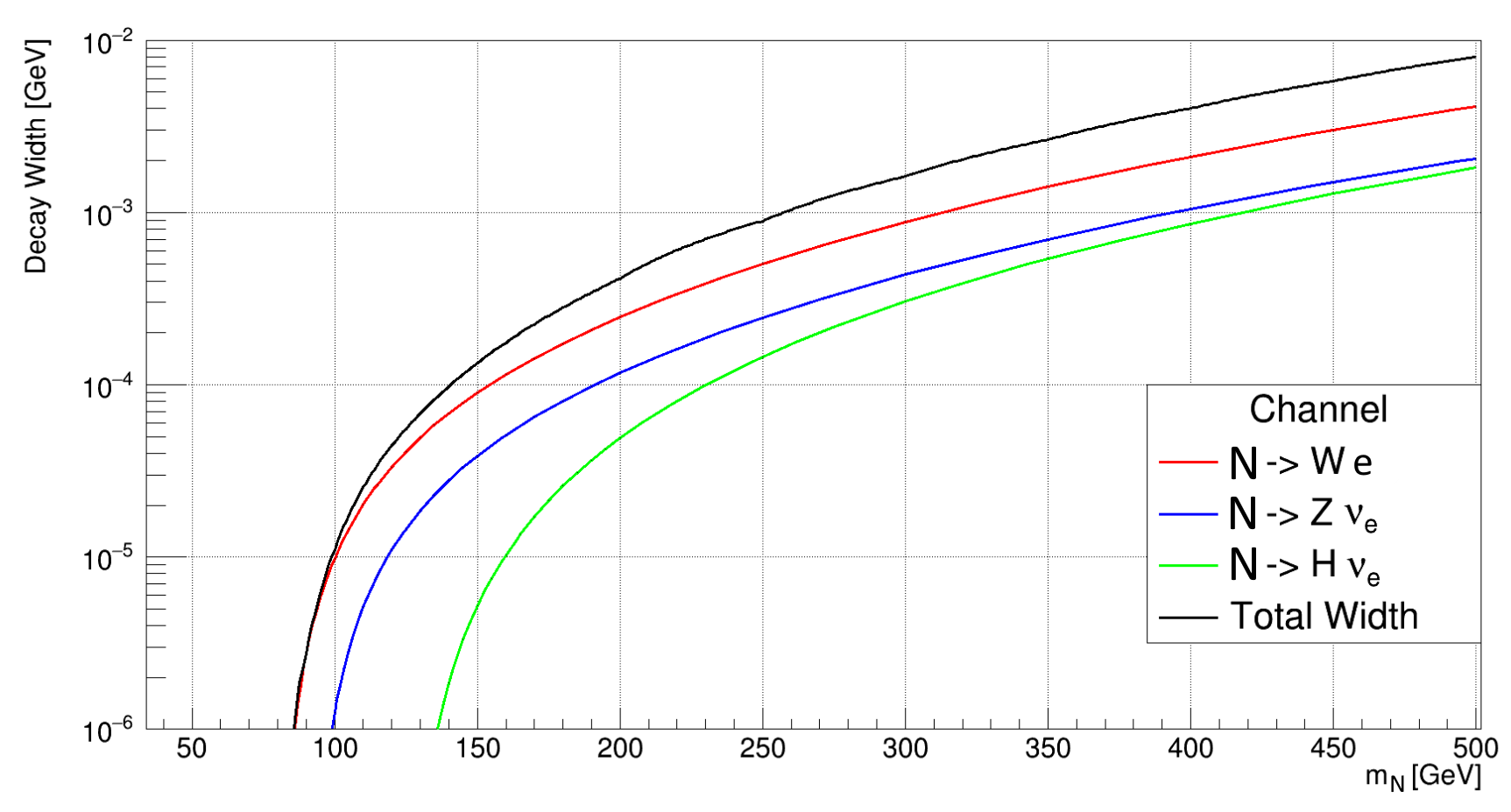}\\
\includegraphics[width=0.8\textwidth]{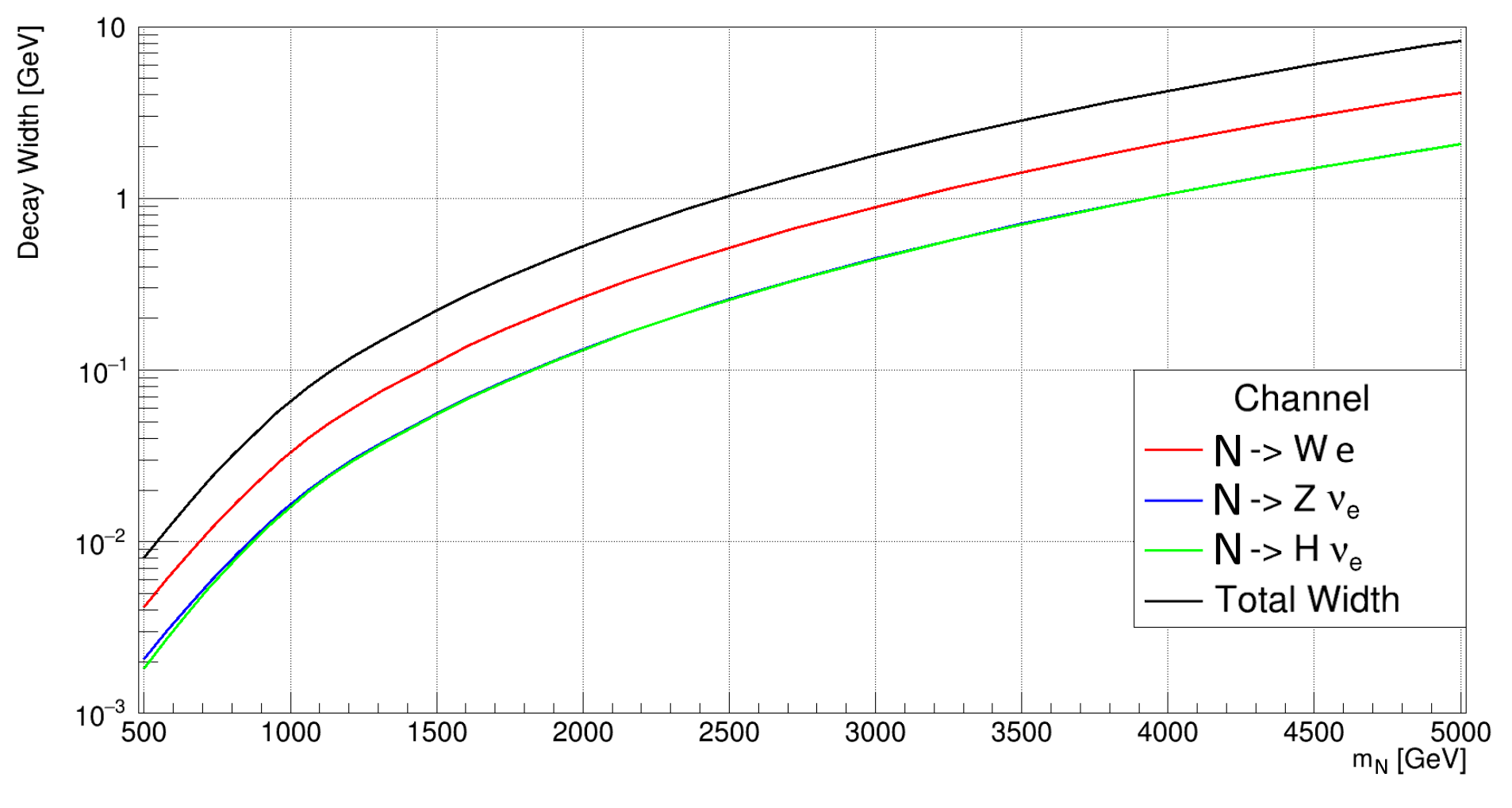}
\caption{Decay width dependence on the iso-doublet neutral VLL mass in the case that $s^E_L = s^N_L$, $s^E_R = 0$ and $s^N_R = 0.01$.}
\end{figure}

In the case of $s^N_R = 0$ only charged current decay mode survived: $\mathrm{BR}(N \to W^+e^-) = 1$. The mass dependence of the decay width coincides with red line in Figure 6 if one assumes that $s^E_L = s^N_L$ and $s^E_R = 0.01$.

In the realistic situation, the mixing angles are expected to be different from each other and from zero. As an example, let us consider the case where right-handed mixings predominate ($s^E_R, s^N_R \gg s^E_L, s^N_L$). In this case, charged current decay mode is proportional to $(s^E_R)^2$ and neutral current decay modes are proportional to $(s^N_R)^2$. In Figure 7, we present dependencies of branching ratios on $s^E_R$ for $m_N = 1000$ GeV and $s^N_R = 0.01$.

\begin{figure}[H]
\centering
\includegraphics[width=0.8\textwidth]{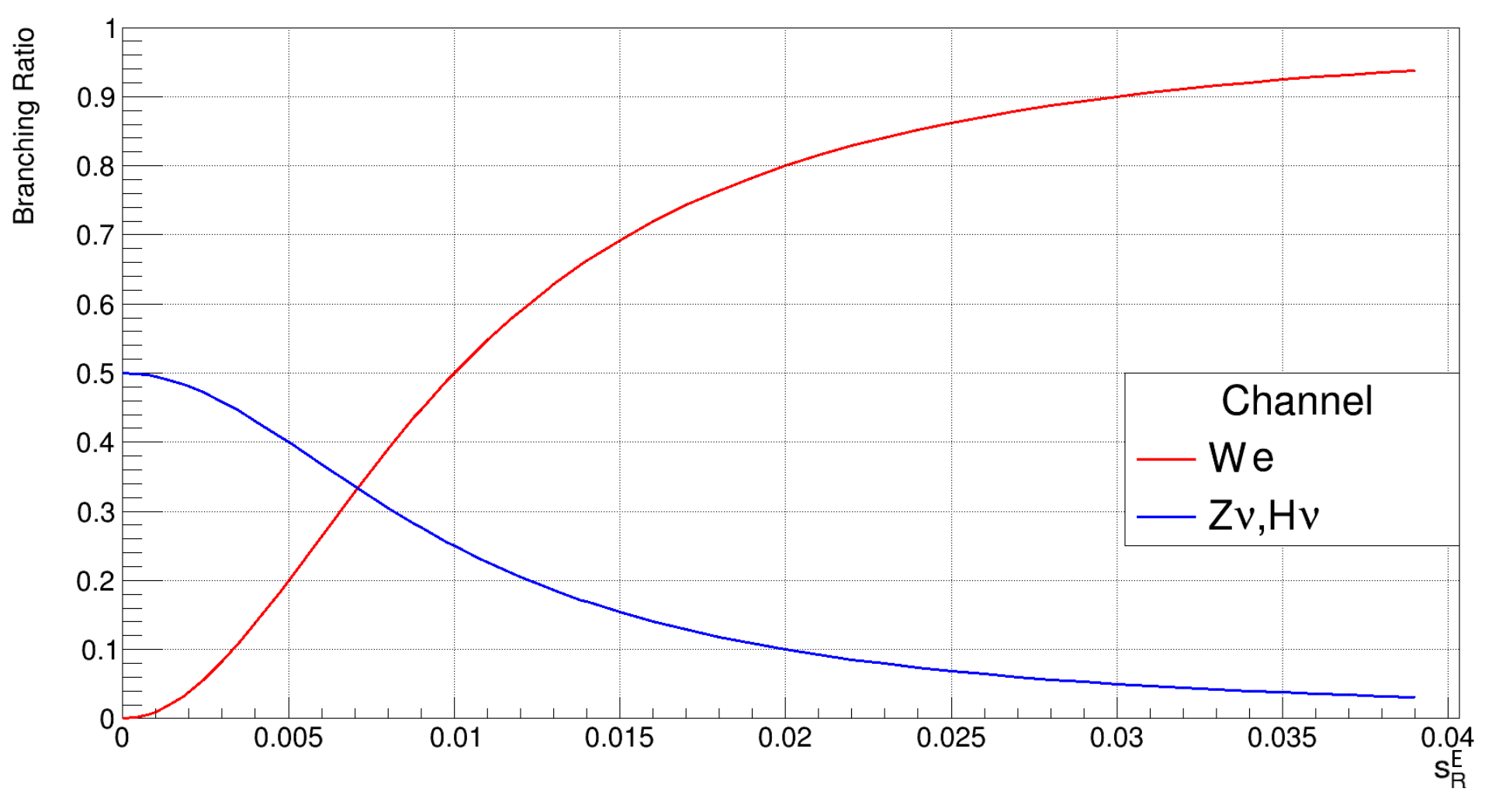}
\caption{Branching ratios dependencies on $s^E_R$ in the case of $s^N_R = 0.01$, $s^E_R$, $s^N_R >> s^E_L$ , $s^N_L$ and $m_N = 1000 GeV$.}
\end{figure}

If left-handed mixings are predominant, the main decay mode is $N \to W^+e^-$: $\mathrm{BR}(N \to W^+e^-) \approx 1$. In this case decay width is proportional to $(c^N_L s^E_L - c^E_L s^N_L)^2$.

\subsection{The case uncovered by current analysis}

Current experimental analyses performed by the ATLAS and CMS collaborations are primarily signature-based and rely on multi-charged lepton final states only. However, in the specific case where $s^E_L = s^N_L$ and $s^E_R = 0$, only the neutral current decay modes survive: $N \to H\nu$ and $N \to Z\nu$ (see Figure 5). A similar situation occurs if $s^N_R$ is dominant ($s^N_R \gg s^E_L, s^N_L, s^E_R$), since the charged current decay mode is severely suppressed.

In this scenario, any charged leptons in the final state would originate exclusively from $Z \to l^+l^-$ decay and $H \to WW^*, W \to l\nu$ chain. Since the branching ratios for this leptonic channels are quite small, their contribution to the multi charged lepton final states are almost negligible. Therefore, the existing analyses remain completely blind to the $N \to H\nu$ and $N \to Z\nu$ channels.

Let us consider the pair production of neutral VLLs that decay exclusively into the $H\nu$ and $Z\nu$ channels. For these final states, we consider the dominant hadronic decay modes: the Higgs boson decaying into a bottom quark pair ($H \to b\bar{b}$), which has a branching ratio of approximately 0.6, and the $Z$ boson decaying into light quarks ($Z \to q\bar{q}$), with a branching ratio of about 0.7.

According to Figure 1, the pair production cross-section for a neutral VLL with a mass of 500 GeV is approximately 6 fb. For an integrated luminosity of 140 fb$^{-1}$, this corresponds to 840 produced events. Taking the relevant branching ratios into account, approximately 140 events would be produced for the specific hadronic final states under consideration.

As a result, the distinct signature of this final state consists of two $b$-jets, two light jets, and large missing transverse momentum ($p_T^{\rm miss}$) originating from the neutrinos. The invariant mass of the $b$-jet pair is expected to reconstruct the Higgs boson mass ($m_H$), whereas the invariant mass of the two light jets will reconstruct the $Z$ boson mass ($m_Z$). If a dedicated search strategy targeting this exact kinematic topology is implemented, the observation of a 500 GeV neutral VLL is virtually guaranteed, making it impossible to miss.

In a recent study \cite{ATLAS:2024higgsino}, pair production of higgsino subsequently decaying to the Higgs boson and gravitino have been considered. Gravitinos lead to large $p_T^{\rm miss}$ and both Higgs bosons are decaying to bb channel. A similar signature occurs in our case if both neutral VLLs are decaying to $H\nu$. The upper limit on the higgsino pair-production cross-section for a branching ratio of $\mathcal{B}(\tilde{H} \to H + \tilde{G}) = 100\%$ is given in Figure 8, which is a cropped version of Figure 12a from \cite{ATLAS:2024higgsino}. It is seen that Higgsino masses below 940 GeV are excluded. It should be noted that in the analysis charged and neutral higgsinos are assumed to be degenerate, which leads to an order of magnitude enhanced cross-section for neutralino pair production.

\begin{figure}[H]
\centering
\includegraphics[width=0.8\textwidth]{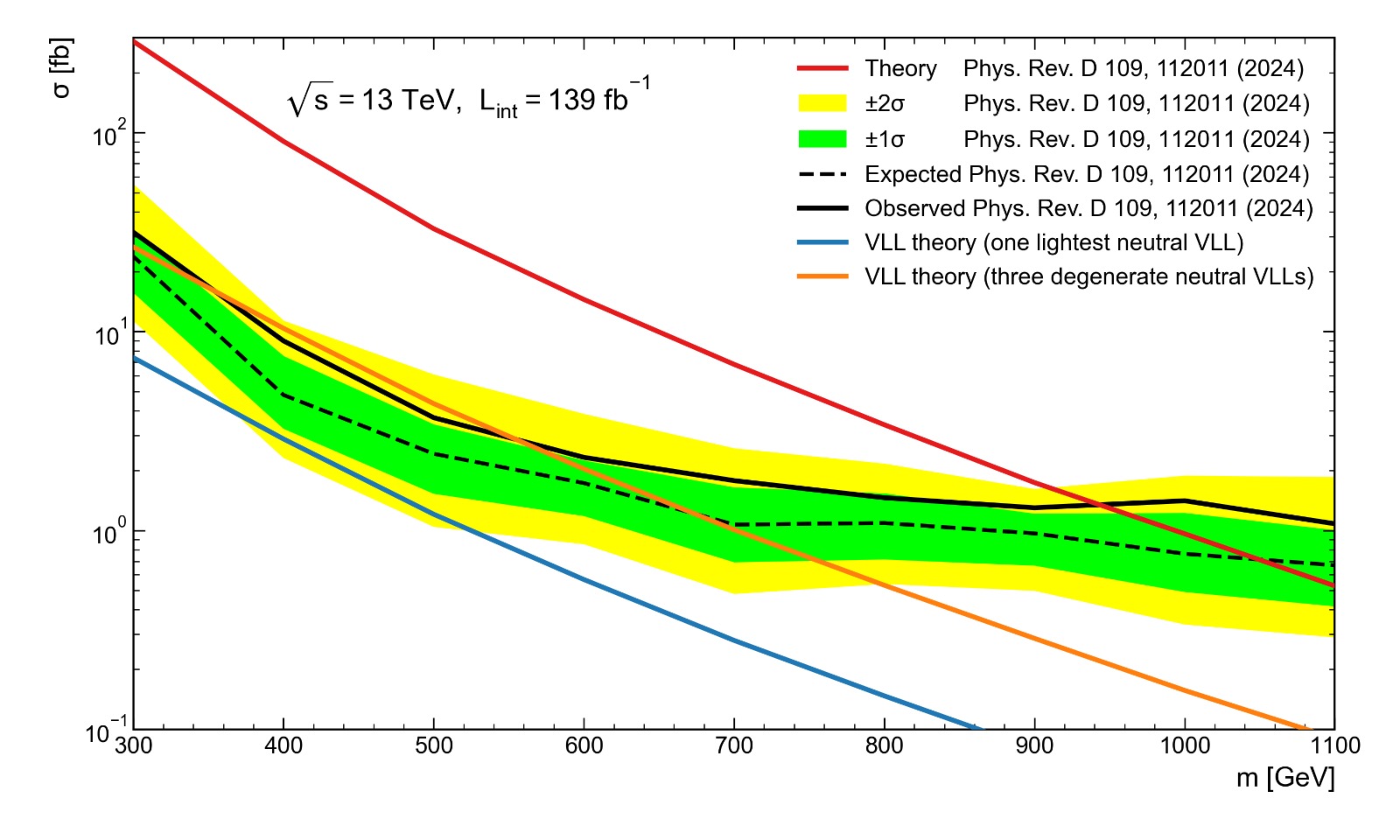}
\caption{The upper limit on the higgsino pair-production cross-section for a branching ratio of $\mathcal{B}(\tilde{H} \to H + \tilde{G}) = 100\%$. The blue line corresponds to the case where one of the neutral VLLs is the lightest and the orange line represents the degenerate state of 3 neutral VLLs.}
\end{figure}

We added the predicted cross-sections from the neutral VLL pair production to Figure 8. The blue line corresponds to the case where one of the neutral VLLs is the lightest and the orange line represents the degenerate state of 3 neutral VLLs. As can be seen, in the first case, the expected cross-section is below the observed limit. In the degenerate case, neutral VLLs with masses between $\approx 355$ and $\approx 555$ GeV are excluded.

Experimental analysis of the case where one of the $Z$ bosons decays to the $\ell^+\ell^-$ ($\ell = e, \mu$) channel does not change these conclusions. For this purpose we use the CMS search in final states with two opposite-charge same-flavor leptons and missing transverse momentum, based on 137 fb$^{-1}$ of data at $\sqrt{s} = 13$ TeV \cite{CMS:2021}. This analysis contains a dedicated ``HZ'' signal region, which requires an on-$Z$ lepton pair ($86 < m_{\ell\ell} < 96$ GeV) and exactly two $b$-tagged jets with $m_{b\bar{b}} < 150$ GeV, and is binned in $p_T^{\rm miss}$. The results are interpreted in a model of pair-produced higgsinos where each $\tilde{\chi}^0_1$ decays to $Z\tilde{G}$ or $H\tilde{G}$ with 50\% probability and $m_{\tilde{G}} = 1$ GeV. This is the same topology as $N\bar{N} \to (Z\nu)(Z\nu),$ $(Z\nu)(H\nu),$ $(H\nu)(H\nu)$ in the case where only neutral current decay modes survive (Figure 5), since $BR(N \to Z\nu) \approx BR(N \to H\nu) \approx 0.5$ for $m_N \gtrsim 400$ GeV. The comparison is shown in Figure 9. It is seen that for the lightest neutral VLL the expected cross-section is still below the observed limit, whereas in the case of three degenerate neutral VLLs masses below $\approx 575$ GeV are excluded, which is somewhat stronger than the bound obtained from Figure 8.

Let us mention that in the case of the restricted model, the ATLAS analysis \cite{ATLAS:2025vll} excludes both neutral and charged VLLs with masses below 1.2 TeV. However, in the case of the more realistic general model, the ATLAS data do not provide any constraint on the lightest neutral VLL!

\begin{figure}[H]
\centering
\includegraphics[width=0.8\textwidth]{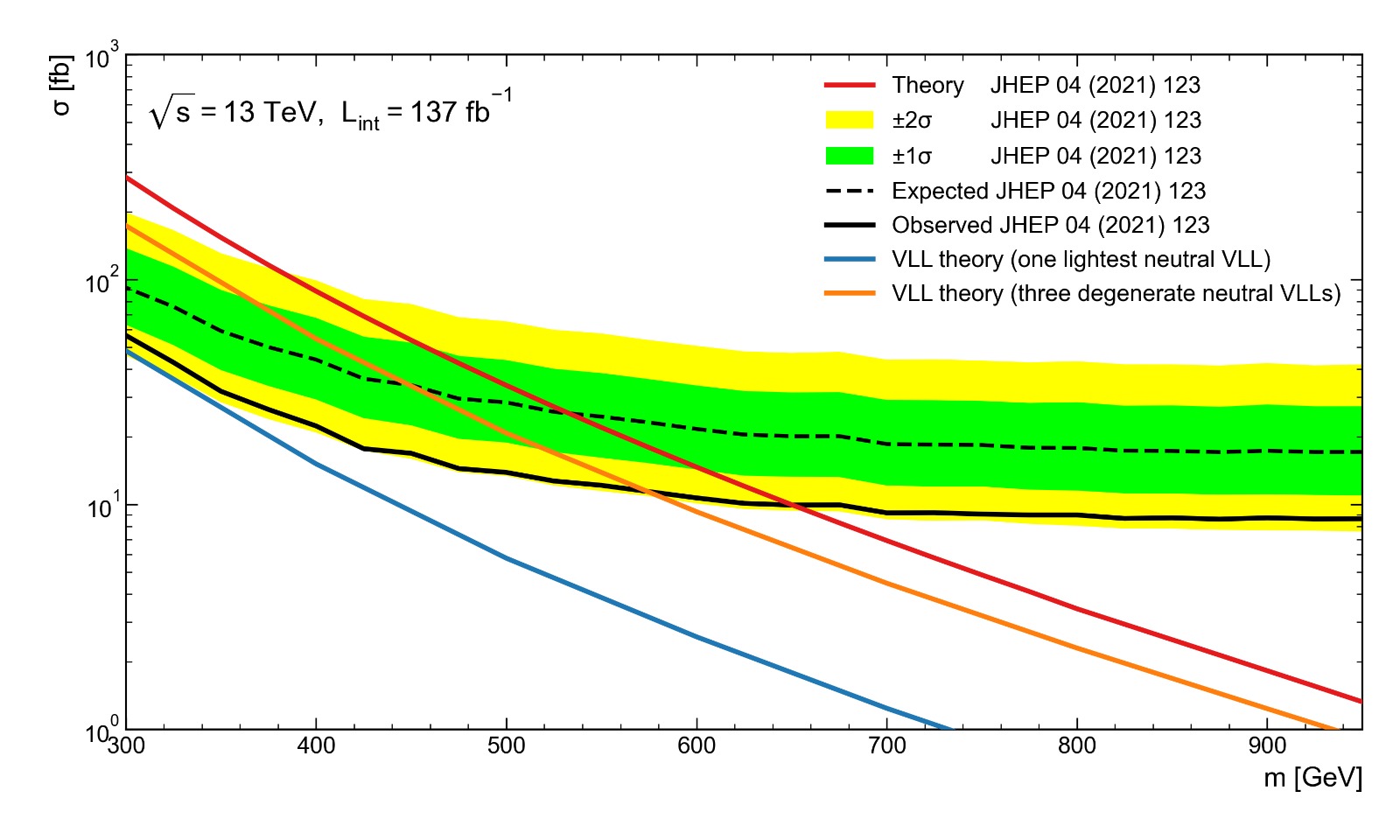}
\caption{Predicted cross-sections for neutral VLL pair production compared with the 95\% CL upper limits of the CMS search in the $\ell^+\ell^-$ + jets + $p_T^{\rm miss}$ final state for $B(\tilde{\chi}^0_1 \to Z\tilde{G}) = B(\tilde{\chi}^0_1 \to H\tilde{G}) = 50\%$ \cite{CMS:2021}. The blue line corresponds to the case where one of the neutral VLLs is the lightest and the orange line represents the degenerate state of 3 neutral VLLs.}
\end{figure}

\section{Conclusion}

There are strong phenomenological arguments for the existence of vector-like leptons, such as Flavor Democracy hypothesis (down-up approach), $E_6$ GUT (up-down approach) and others, and corresponding experimental studies should be strengthened. Let us emphasize that (as mentioned in \cite{Baspehlivan:2022}) ``Discovery of VLLs and VLQs will shed light on the mass and mixing patterns of fundamental (at today's level) fermions.'' As pointed out in the Introduction, several years ago Steve Weinberg stated that the observed pattern of quark and lepton masses is one of the most important mysteries to be solved \cite{Chalmers:2017}.

A number of possible decay channels of charged and neutral iso-doublet vector-like leptons have been omitted in experimental analyses carried out so far. Two reasons for disregarding these channels are the degenerate masses of charged and neutral VLLs and the assumption of the absence of right-handed neutrinos in the restricted model. While the former may be true if only one VLL pair is present, neutrino oscillations disprove the latter.

Let us remind that:

i) The assumption of degeneracy is valid if only one vector-like lepton doublet exists. In much more natural case each SM family should have its own VLL doublet, as in $E_6$ GUT. Therefore, all production channels (pair production of neutral VLLs, pair production of charged VLLs and associate production of charged and neutral VLLs) should be analyzed separately,

ii) Decay channels induced by the existence of $\nu_R$ should be taken into account,

iii) The unsuppressed decay channel $E \to NW$, which is dominant if $m_E > m_N$, should not be ignored.

Therefore, it would be beneficial to update the ATLAS and CMS analyses by including the missing channels. Hopefully the HL-LHC studies will also take this into account.

Finally, the results of Section 4.3 show that the lightest neutral VLL decaying only via neutral currents ($N \to Z\nu$ and $N \to H\nu$) is not excluded by the ATLAS \cite{ATLAS:2024higgsino} and CMS \cite{CMS:2021} data: its predicted cross-section remains below the observed limits both in the $HH + p_T^{\rm miss}$ and in the $\ell^+\ell^-$ + jets + $p_T^{\rm miss}$ final states. In the case of three degenerate neutral VLLs, masses below 555 GeV are excluded in the $HH$ channel and masses below 575 GeV in the $\ell^+\ell^-$ channel.

\section*{Acknowledgements}

We are grateful to Ali Can Canbay, \"Umit Kaya and Arda Goroglu for useful discussions.

\end{document}